\documentclass[twocolumn, showpacs,preprintnumbers,amsmath,amssymb]{revtex4-1}

\usepackage{graphicx}% Include figure files
\usepackage{dcolumn}% Align table columns on decimal point
\usepackage{bm}% bold math
\newcommand{\balpha}{\boldsymbol{\alpha}}

\begin{document}
 \preprint{APS/123-QED}

\title{Path Integral Formulation of Anomalous Diffusion Processes }% Force line breaks with \\

%\date{\today}% It is always \today, today,
\author{R.~Friedrich}
\affiliation{Institute for Theoretical Physics, University of M\"unster, Wilhelm-Klemm-Str. 9, 48149 M\"unster, Germany}
\author{S.~Eule}
\affiliation{Max Planck Institute for Dynamics and Self-Organization, Am Fa{\ss}berg 17, 37077 G\"ottingen, Germany\\}

\date{\today}

\begin{abstract}
We present the path integral formulation of a broad class of generalized diffusion processes.
Employing the path integral we derive exact expressions for the path probability densities and joint
probability distributions for the class of processes under consideration.  We show that 
Continuous Time Random Walks (CTRWs) are included in our framework. A closed expression for the path probability
distribution of CTRWs is found in terms of their waiting time distribution as the solution of a Dyson equation. As the formalism
naturally includes the treatment of functionals a generalized Feynman-Kac formula is derived.
\end{abstract}

% PACS, the Physics and Astronomy
                             % Classification Scheme
%\keywords{Suggested keywords}%Use showkeys class option if keyword
                              %display desired
\maketitle

{\em Introduction}:
In addition to Langevin equations and Fokker-Planck equations, 
the concept of path integrals plays a central role in the description of
stochastic processes \cite{Langouche, Kleinert}.
The path integral formulation is based on the specification of a probability
measure assigned to each realization of the process and, hence, 
encodes the complete statistical information on the process. 
It was pioneered
by Onsager and Machlup who derived an expression for the probability density of a path of a linear
Gaussian process \cite{Onsager}. 
Later their work has been extended to include nonlinear drift and
diffusion coefficients as well as coloured noise \cite{Haken}.

Despite the success of models based on Brownian motion and diffusion processes,
over the last two decades it has become apparent that many dynamical systems in 
diverse fields, ranging from biology to physics, cannot aptly be described 
within this framework \cite{Bouchaud, Metzler, Shlesinger}.  
Various generalizations of diffusion processes have been considered to account 
for such anomalous diffusion. Some of
these generalized models consist of an additional stochastic process. 
Examples include the Continuous Time Random Walk (CTRW) \cite{MoWe}
which has been applied to model a variety of different systems \cite{Metzler}.
The CTRW is defined as a proper random walk with an additional 
random process governing the waiting time between successive jumps.
If the waiting time distribution is assumed to be scale-free the CTRW can
be described by the celebrated fractional Fokker-Planck equation (For a
review, we refer the reader to \cite{Metzler}).
The corresponding Langevin equation can be formulated by application of the 
concept of subordination \cite{Fogedby}. A path integral
formulation of CTRWs has not been established so far. Such a formulation, 
however, is of particular importance since CTRWs in general are
non-Markovian processes and as such insufficiently described by 
single-point probability distributions \cite{Baule}. 
Another example where stochastic dynamical processes are combined with an 
additional random process is the field of superstatistics \cite{Beck}. 

This Letter presents a path integral approach to diffusion processes 
which are additionally driven by a further
stochastic process. This class of processes comprises the CTRW and 
thus fills the gap of the lacking path integral formulation for CTRW's.

{\em Path integral representation}:
The class of processes under consideration in this Letter is defined by the one-dimensional discrete Langevin equation
\begin{eqnarray}\label{Langevin2}
q_{k+1}=q_k+\tau N(q_k)+\sqrt{\tau D}R_k+\alpha_k r_k\, ,
\end{eqnarray}
where the first three terms on the right hand side describe a standard Langevin equation with time step $\tau$, drift $N(q_k)$ and 
Gaussian random variables $R_k$ with amplitude $\sqrt{D}$. The anomalous 
contribution stems from the last term where 
$\alpha_k$ and $r_k$ both are random variables. The transition amplitude
of this process can be obtained as
\begin{eqnarray}\label{transition}
\lefteqn{
p(q_{k+1} |q_{k},\alpha_{k})=}
\nonumber \\
&&
\int \frac{d \tilde{q_k}}{2\pi}
e^{i \tilde q_k (q_{k+1}-q_k-\tau N(q_k)-\alpha_k r_k)-\frac{\tau D}{2} \tilde q_k^2} \, ,
\end{eqnarray}
where we assumed the stochastic force $R_k$ to be Gaussian 
distributed with zero-mean
and vanishing correlation $\langle R_k R_l\rangle=  \delta_{jk}$. 
The probability for a path
starting at $q_0$ is given by iteration of Eq.(\ref{transition})
\begin{eqnarray}
\lefteqn{
g_{k+1}(q_{k+1}; q_k,\alpha_{k}; ..., q_0, \alpha_0)=}
\\
&&
\int \mathcal{D}{\tilde q} e^{S_0(q, {\tilde q})} e^{-i\sum_{k=0}^{N} {\tilde q_k} \alpha_k r_k} ,
\nonumber 
\end{eqnarray}
where $\mathcal{D}{\tilde q}= \prod_{k=0}^N \frac{d{\tilde q}_k}{2\pi}$ and $S_0(q,\tilde q)$ is the 
Martin-Siggia-Rose action of a diffusion process with drift \cite{Shraiman}, 
written in discretized form, 
\begin{equation}\label{MSR}
S_0(q,\tilde q)=i \sum_{k=0}^{N} \lbrace
\tilde q_k (q_{k+1}-q_k-\tau N(q_k))+i \frac{\tau D}{2} \tilde q_k^2 \rbrace \, .
\end{equation} 
To further evaluate the expression for the path probability, let us assume that 
the $r_k$ are Gaussian random variables with zero mean and variance 
$\langle r_k r_l \rangle =\tau Q\delta_{kl}$. After averaging with respect 
to the $r_k$ we find
\begin{eqnarray}
\lefteqn{
g_{N+1}(q_{N+1};q_{N},\alpha_{N};..;q_0,\alpha_0)=}
\nonumber \\
&&
\int \mathcal{D}\tilde q
e^{i \sum_{k=0}^{N} \lbrace \tilde q_k (q_{k+1}-q_k-\tau N(q_k))+\frac{i\tau (D+Q\alpha_k^2)}{2} \tilde q_k^2 \rbrace}
\nonumber \\
&&
=\frac{1}{\sqrt{\prod_{k=0}^{N}2\pi\tau (D+Q\alpha_k^2)}}
e^{-\sum_{k=0}^{N} \frac{(q_{k+1}-q_k-N(q_k))^2}{2\tau(D+Q\alpha_k^2)}} \, .
\end{eqnarray}
The probability density for a specific path $f_{N+1}(q_{N+1};..;q_0)$ 
is then given by averaging with respect to the stochastic process $\alpha_k$
\begin{eqnarray}\label{pathprob}
\lefteqn{
f_{N+1}(q_{N+1};..;q_0)
= }
\\
&&
\int \mathcal{D} \tilde q
e^{S_0(q,\tilde q)} Z(i\frac{\tau Q}{2} \tilde q_N^2,..,i\frac{\tau
  Q}{2}\tilde q_0^2) \nonumber 
 \, 
\end{eqnarray}
We have introduced the characteristic function
\begin{eqnarray}\label{pathprob2}
Z(\eta_N,..,\eta_0) & = & \langle e^{i \sum_{k=0}^N \eta_k \beta_k}\rangle \nonumber \\
 & = & \sum_{\alpha_N} ..\sum_{\alpha_0} p(\alpha_N;..;\alpha_0)
e^{i \sum_{k=0}^N \eta_k \alpha_k^2} \, .
\end{eqnarray}
which is the characteristic function of the process $\beta_i=\alpha_i^2$. 
We have denoted the distribution of $\balpha=[\alpha_N,..,\alpha_0]$
by $p(\balpha)$.
\\
{\em Propagators}:
The representation (\ref{pathprob}) for the path probability serves as the starting point to determine
propagators for the anomalous stochastic process defined by Eq.(\ref{Langevin2}). As usual,
the propagator is obtained by integration of all possible paths with appropriate boundary conditions
\begin{equation}\label{pathintegral}
\mathcal{G}(q_{N}; q_0, N)=
\int \mathcal{D}q f_N(q_N,..,q_1,q_0) \, .
\end{equation}
%\int_{q_0}^{q_N} \mathcal{D}q \int \mathcal{D} \tilde q 
%e^{S_0(q,\tilde q)} \sum_{\balpha}
%p(\balpha)e^{-\tau \frac{Q}{2}\sum_{k=0}^{N-1} \alpha_k^2 \tilde q_k^2}.
%\end{equation}
Let us consider a process consisting of two pure diffusions, i.e. we set
$N(q_k)=0$. We obtain
\begin{equation}\label{example}
\mathcal{G}(q;q_0, N)=\sum_{\balpha} p(\balpha) 
\frac{e^{-\frac{(q-q_0)^2}{2\tau\sum_{k=0}^{N-1}[D+Q \alpha^2_k]}}}
{\sqrt{2\pi\tau(\sum_{k=0}^{N-1}[D+Q \alpha^2_k])}} \, .
\end{equation}
Observe that the quantity $S(N)=\sum_{k=0}^{N-1} \alpha_k^2$ plays a key role for the diffusion
process. After introduction of the pdf $P_N(S)=\sum_{\balpha} p(\balpha) \delta_{S,\sum_{k=0}^{N-1} \alpha_k^2}$ 
we can rewrite Eq.(\ref{example}) as
\begin{equation}
\mathcal{G}(q_N, q_0, N)=\sum_{S=0}^\infty P_N(S) 
\frac{e^{-\frac{(q-q_0)^2}{2\tau(ND+SQ)}}}
{\sqrt{2\pi\tau(ND+SQ)}}\,.
\end{equation}
In comparison to the regular diffusion process which takes place in the physical time $N\tau$ 
the second diffusion process can be interpreted to occur in the random time $S\tau $ whose distribution is determined
by $p(\balpha)$. A corresponding result for a continuous jump-diffusion process was obtained in \cite{Eule1} with
different methods.

%S denotes the number of events,i.e. the number of ones in the string, occuring
%$ \rightarrow Q\tau$, $t=N\tau$, and introduce the probability distribution
%$P_N(S) \rightarrow p(s,t)$. Then
%\begin{equation}
%f(q,t)=\int ds p(s,t) 
%\frac{e^{-\frac{(q-q_0)^2}{2(Qt+s)}}}
%{\sqrt{2\pi(t Q+s)}}
%\end{equation}

Multiple-time propagators can be obtained from the 
propagators $\mathcal{G}(q_N, q_0, N)$ in a straightforward manner
\begin{eqnarray}
\mathcal{G}(q_N,q_M,q_0) &=& \sum_{S_N=0}^\infty\sum_{S_M=0}^\infty  P_{N-M,
  M}(S_N-S_M;S_M) \nonumber 
\\
 &\times& \mathcal{G}(q_N;q_M, N-M) \mathcal{G}(q_M;q_0, M).
\end{eqnarray}
where  
$P_{N-M, M}(S_N-S_M;S_M)$ 
denotes the probability distribution of the variables
$S_N-S_M=\sum_{k=M}^{N-1} \alpha_k^2$ and $S_M=\sum_{k=0}^{M-1} \alpha_k^2$, which
can be evaluated using the probability distribution $p(\balpha)$.
The generalization to $n$-time pdfs is straightforward and
is given in terms of the pdf $P_{N-k, k-l, ..., m}(S_N-S_{k}; S_{k}-S_{l};..S_m)$ and the
product of the propagators $\mathcal{G}(q_N;q_M, M-N)$.

{\em Examples for the $\balpha$-process}:
All results obtained so far hold for a general process $\balpha$. 
In the following we consider
two specific examples for this process. 

The case where the $\alpha_k$ are 
i.i.d. random variables with a common pdf $h(\alpha)$ is in the realm of 
superstatisics \cite{Beck}.

%For example if $\beta=\alpha^2$ is distributed according to an inverse  $\chi^2$ distribution, it can be shown
%that the pdf $f(q)$  is q-Gaussian \cite{Beck}. 

Renewal processes are obtained when the
process $\alpha_k$ is a binary string. If we denote the number of zeros 
between two successive ones by $t$ and assume them to be i.i.d. random
variables with a common pdf $W(t)$, the process $\balpha$ is a
renewal process. In this case the process described by 
Eq.(\ref{Langevin2}) is a CTRW with internal dynamics 
\cite {Eule1, Friedrich} which for $N(q_k)=D=0$ includes the standard CTRW.

{\em  Path integrals for CTRWs:}
The starting point of our treatment of CTRW path integrals are
Eqs. (\ref{pathprob}) and (\ref{pathprob2}) and we need to characterize 
the process $\balpha$. For the sake of simplicity, we restrict ourselves
to processes with $\alpha_0=1$, i.e. processes that start with an event. 
The aim is to represent the probability
of a specific sequence $p(\balpha)$ in terms of the waiting time distribution, i.e. the number of zeros between two successive ones.
Let $W_{i,j}$ denote the probability to have $i-j-1$ zeros between ones at $j$
and $i$, $i > j$ and let furthermore $W_{i,i}=0$.
It follows that the survival probability, 
i.e. the probability that
a one at $j$ is not  followed by a further one till $i$ is $1-W_{i,j}$.

To proceed it is convenient to introduce the probability 
$\nu_k(1, \alpha_{k-1}, ...,\alpha_1,1)$ of
truncated strings, $[1,0,0,..,1,..,0,1]$
which end with the event at $k$. 
According to the definition of renewal processes
the truncated densities fulfill the relation 
\begin{eqnarray}\label{renewal}
\lefteqn{
\nu_k(1, \alpha_{k-1}, ...., \alpha_1,1)=\delta_{k,0}}
\nonumber \\
&&
+\sum_{l=0}^{k} \sigma_{kl}
\nu_l(1, \alpha_{l-1}, ...., \alpha_1, 1) \, ,
\end{eqnarray}
where we have defined 
\begin{equation}
\sigma_{kl}= 
\delta_{\alpha_k,1}\delta_{\alpha_{k-1},0}..\delta_{\alpha_{l+1},0}\delta_{\alpha_l,1}
W_{k,l} \, .
\end{equation}
The density of the truncated strings for $k=N$ depends on the densities
for $k<N$. Iterative application of (\ref{renewal}) allows us to
present the truncated density for $k=N$ according to 
\begin{eqnarray}
\lefteqn{
\nu_N(1, \alpha_{k-1}, ...., \alpha_1,1)=}
\\
&&
\sigma_{N0}+\sum_{l=0}^N
\sigma_{Nl} \sigma_{l0}+ \sum_{l=0}^N\sum_{l'=0}^l \sigma_{Nl}\sigma_{ll'}\sigma_{l'0} + ... \nonumber \\ 
&& ...+ \sigma_{NN-1}\sigma_{N-1N-2} ...\sigma_{21}\sigma_{10} \, ,
\nonumber 
\end{eqnarray}
where $N>l>l'> ... >0$.
This series can be summed, yielding
\begin{equation} \label{summed}
\nu_k(1, \alpha_{k-1}, ...., \alpha_1,1)=
\sum_{l=0}^N [E-\boldsymbol{\sigma}]^{(-1)}_{kl} \sigma_{l0}\, ,
\end{equation}
where $E$ denotes the unit matrix and $\boldsymbol{\sigma}$ is a matrix with elements $\sigma_{ij}$. 
Relation (\ref{summed}) is valid for $k\le N$ and implies that the truncated
densities are defined by a Dyson equation 
whose explicit solution is given by matrix inversion of
$E-\boldsymbol{\sigma}$.

The probability density 
$p(\alpha_N,...,\alpha_0)$ of the renewal process then is determined
from the truncated densities on the basis of the relationship
\begin{equation}\label{renewal2}
p(\alpha_N,...,\alpha_0)=\sum_{l=0}^N \gamma_{Nl}
\nu_l(1,\alpha_{l-1},..,\alpha_1,1) \, ,
\end{equation}
where we have defined 
\begin{equation}
\gamma_{Nl}=\delta_{\alpha_N,1}\delta_{N,l}+(1-W_{Nl})\delta_{\alpha_N,0}\delta_{\alpha_{N-1},0}..
\delta_{\alpha_{l+1},0} \delta_{\alpha_l,1} \, .
\end{equation}
The explicit representation of the probability density is then just
\begin{equation}\label{renewal22}
p(\alpha_N,...,\alpha_0)=\sum_{k=0}^N\sum_{l=0}^N
 \gamma_{Nk} [E-\boldsymbol{\sigma}]^{(-1)}_{kl} \sigma_{l0} \, .
\end{equation}
Again, an expansion of the matrix $[E-\boldsymbol{\sigma}]^{-1}$ yields a representation of
the probability density of a string $[0,..,1,..1,..1]$ in the number of one's
contained in the string.
Concluding, we state the characteristic function of the renewal process
$\balpha$ which can be easily assessed on basis of Eq.(\ref{renewal22}):
\begin{equation}\label{charac}
Z(\eta_N,..,\eta_0)=
\sum_{k=0}^N\sum_{l=0}^N
 \tilde \gamma_{Nk} [E-\tilde{\boldsymbol{\sigma}}]^{(-1)}_{kl} \tilde \sigma_{l0} \, ,
\end{equation}
where now
\begin{equation}
\tilde \gamma_{kl}=(1-W_{kl}) e^{i \eta_l} \qquad , \qquad 
\tilde \sigma_{kl}=e^{i \eta_k} W_{kl} e^{i \eta_l} \, .
\end{equation}
The combination of Eq.(\ref{pathprob2}) with Eq.(\ref{charac}) provides the 
desired path-integral representation for the class of CTRWs under consideration.   

%Explicitly, we obtain the representation of $p(\balpha)$
%\begin{eqnarray}\label{probdis1}
%\lefteqn{
%p(\alpha_N,...,\alpha_0)=(1-W_{N,0})\delta_{\alpha_N,0}...\delta_{\alpha_0,1}
%}
%\nonumber \\
%&&+\sum_{l=1}^N (1-W_{N,l})W_{l,0} \delta_{\alpha_N,0}..\delta_{\alpha_l,1}..
%\delta_{\alpha_0,1}
%\nonumber \\
%&&+\sum_{l=1}^N (1-W_{N,l})\sum_{l'=1}^l W_{l,l'} W_{l',0}
%\delta_{\alpha_N,0}..\delta_{\alpha_l,1}..\delta_{\alpha_{l'},1}..
%\delta_{\alpha_0,1}
%\nonumber \\
%&& + ... \nonumber \\
%&&+ W_{N,N-1}W_{N-1,N-2} ..W_{1,0}
%\delta_{\alpha_N,1}...\delta_{\alpha_0,1} \, ,
%\end{eqnarray}
%This is an expansion in terms of the numbers of events 1 occuring in the
%string characterizing the renewal process. 

Since this path-integral representation is a rather condensed representation
of the CTRW we proceed to derive a more transparent
formulation. It is based on Eqs. 
(\ref{renewal}) and (\ref{renewal22}). It is clear, that the probability
distribution $f_{N+1}(q_{N+1},q_N,..q_0)$ should have an analogous representation.
We introduce the abbreviations
$K_0(q_N; ...;q_1)$ for the path probability of the MSR action (\ref{MSR}) and 
 $V(q_{l+1};q_l)$ for the short time propagator of the action with $\alpha_l=1$
\begin{equation}
V(q_{l+1};q_l)=\frac{1}{\sqrt{2\pi \tau (D+Q)}} e^{-\frac{(q_{l+1}-q_l-N(q_l))^2}{2\tau(D+Q)}}\, .
\end{equation}
The starting point is the definition of the truncated distribution
\begin{eqnarray}
\lefteqn{
\eta_N(q_{N+1},q_N,...,q_0)=
\sum_{\alpha}
\nu_N(1,\alpha_{N-1},..,\alpha_1,1)
}
\\
&&
~~~~~~~~~\times~~g_{N+1}(q_{N+1};q_N,1;q_{N-1},\alpha_{N-1};..;q_0,1) \, .
\nonumber 
\end{eqnarray}
Using the relation (\ref{renewal}) we arrive at the Dyson equation for
the truncated distribution
\begin{eqnarray}\label{Dysoneq}
\lefteqn{
\eta_k(q_{k+1},q_k,..,q_0)=\delta_{k,0} V(q_1,q_0)}
\\
&&
+\sum_{l=0}^{k} V(q_{k+1},q_k)W_{kl} K_0(q_k,..q_{l+1})\eta_l(q_{l+1},..q_0) \, .
\nonumber 
\end{eqnarray}
Employing Eq.(\ref{renewal2}) we can establish the connection between 
the distributions $\eta$ and $f$:
\begin{eqnarray}\label{dys}
\lefteqn{
f_{N+1}(q_{N+1},q_N,..,q_0)=}
\\
&&
\sum_{l=0}^{N}
(1-W_{Nl}) K_0(q_{N+1},..,q_{l+1})\eta_l(q_{l+1},..q_0) \, .
\nonumber 
\end{eqnarray}
The relations (\ref{Dysoneq}) and (\ref{dys}) yields a representation of
the path-probability of the CTRW in terms of  $K_0$ and $V$. 
As we will show below, this representation can be the starting for determining
expectation values of the path.

It is possible to combine both equations, (\ref{Dysoneq}) and (\ref{dys})
to obtain the representation of the joint probability distribution
as 
\begin{eqnarray}\label{rep}
\lefteqn{
f_{N+1}(q_{N+1},q_N,..,q_0)=
\sum_{l=0}^{N}
(1-W_{Nl}) K_0(q_{N+1},..,q_{l+1})
}
\\
&&
[\delta_{lk} -
V(q_{l+1},q_l)W_{lk} K_0(q_{l},..q_{k+1})]^{(-1)} 
V(q_1,q_0) \delta_{k,0} \, .
\nonumber 
\end{eqnarray}
Thereby, the Dyson equation (\ref{Dysoneq}) was solved explicitly.

%In matrix form
%\begin{eqnarray}\label{Dysoneq}
%\sum_{l=0}^N [\delta_{kl} &-&
%V(q_{k+1},q_k)W_{kl} K_0(q_N,..q_{l+1})]\eta_l(q_{l+1},..q_0)
%\nonumber \\
%&=& \delta_{k,0} V(q_1,q_0)
%\end{eqnarray}

Expansion of the inverse matrix then yields an expansion of the 
path probability in terms of the number
of events. The first term $f^{(0)}$ yields the contribution of paths 
during which no event occurred after the first. 
\begin{equation}
f_{N+1}^{(0)}(q_{N+1}; ...;q_0)=(1-W_{N,0}) K_0(q_{N+1}; ...; q_1)V(q_1;q_0) \, ,
\end{equation} 
whereas the first order term contains 
the contribution from paths with one event 
\begin{eqnarray}
\lefteqn{ 
f_{N+1}^{(1)}(q_{n+1}; ...; q_0) 
=\sum_{l=1}^{N}(1-W_{N,l}) 
\times }
\\
&&
K_0(q_{N+1}; ...; q_{l+1}) V(q_{l+1};q_l)
 W_{l,0} K_0(q_{l}; ...; q_1)V(q_1;q_0) \nonumber \, .
\end{eqnarray}
Higher order terms are obtained in the same way.
Note the formal analogy to the self-energy corrections in quantum theory.
In this case the corrections to the free propagator are given by the anomalous
contributions from the $\balpha$-process. By considering processes which start with an event, 
we imply non-equilibrium conditions.
Relaxing this condition would lead to the case of aging CTRWs \cite{Barkai}.

{\em Generalized Fokker-Planck Equation and Generalized 
Feynman-Kac Formula}: 
We want to conclude with the derivation of the 
generalized Fokker-Planck equation for the CTRW-case. 
Apparently, we can also derive Feynman-Kac formulas \cite{Majumdar}. 
Let us consider the functional
\begin{eqnarray}
\lefteqn{
P_{N+1}(p, q_{N+1})= 
}
\\
&&
\int \prod_{j=0}^{N}dq_j e^{i p \sum_{k=0}^{N} U(q_k)} 
f_{N+1}(q_{N+1},..,q_0) \, .
\nonumber 
\end{eqnarray}
The quantity $P_{N+1}(0,q_{N+1})$ is just the probability distribution
$f_{N+1}(q_{N+1})$ which obeys a generalized Fokker-Planck equation.
Defining now
\begin{eqnarray}
\lefteqn{
G(q_{N+1}, q_l;p, N,l)=}
\\
&&
\int \prod_{j=l+1}^Ndq_j e^{i p \sum_{k=l}^N U(q_k)} f_{N+1-l}(q_{N+1},..,q_l)
\, ,
\nonumber 
\end{eqnarray}
as well as 
\begin{equation}
\zeta_l(p, q_{l+1})=\int \prod_{j=0}^{l+1} dq_j e^{ip\sum_{k=0}^{l}U(q_k)} 
 \eta_l(q_{l+1},..,q_0)\, ,
\end{equation}
we obtain from from Eqs. (\ref{Dysoneq}) and (\ref{dys})  
the relations
\begin{eqnarray}\label{eta1}
\lefteqn{
P_{N+1}(p, q_{N+1}) = \sum_{k=0}^N (1-W_{N,k}) }
 \\
&& \times \int dq_{k+1} G(q_{N+1},q_{k+1};p,N,k)\zeta_l(p,q_{k+1}) \, ,
\nonumber 
\end{eqnarray}

\begin{eqnarray}\label{eta2}
\zeta_k(p, q_{k+1})= V(q_1,q_0)\delta_{l,0}
+\sum_{l=0}^k  \int dq_k
\int dq_{l+1}
\\
 V(q_{k+1},q_k) W_{k,l}
  G(q_k,q_{l+1},k,l+1)\zeta_l(p,q_{l+1})
  \, .
\nonumber
\end{eqnarray}
Now Eqs. (\ref{eta1}) and (\ref{eta2})
are just the time-discrete versions of the equations which served as 
the starting point for the derivation of generalized
Fokker-Planck equation for the CTRW with internal dynamics \cite{Eule1}
eventually leading to 
\begin{equation}\label{genfeynmankac}
\left[\frac{\partial}{\partial t} -\mathcal{H}\right]P(p, q, t)=\int_0^tdt' Q(t-t') \mathcal{L}_1 e^{(t-t')\mathcal{H}} P(p, q, t')\, ,
\end{equation}  
where $\mathcal{H}=\mathcal{L}_0+ip U(q)$. $\mathcal{L}_0$ is the generator
of the Fokker-Planck process with $\alpha=0$  
and $Q(t-t')$ is the common time-evolution kernel of CTRWs \cite{Metzler}.
For the case of CTRWs without 
internal dynamics, i.e. $\mathcal{L}_0=0$, this equation was recently derived in \cite{Turgeman}.

{\em Conclusions and Outlook}
In summary, we have presented the path integral formulation for a class of generalized diffusion processes which
includes the CTRW. Applying the path integral, we have derived general expressions for the propagators and the multipoint
distributions. For the case of  CTRW-processes we found a closed form expression for the probability density of a path in terms of the waiting time
distribution. We expect that this expression will help to evaluate
interesting functionals for anomalous diffusion processes. Furthermore, we
want to mention that it is tempting to consider the case, where the
renewal process $\balpha$ is generated by a one-dimensional map 
\cite{Geisel}, along similar lines.

\end{document}